# The Conducting Channel at the LaAlO$_3$/SrTiO$_3$ Interface

Z. Huang,[1] X. Renshaw Wang,[1,2] Z. Q. Liu,[1,2] W. M. Lü,[1,3] S. W. Zeng,[1,2] A. Annadi,[1,2] W. L. Tan,[2] X. P. Qiu,[3] Y. L. Zhao,[1,2] M. Salluzzo,[4] J. M. D. Coey,[1,5], T. Venkatesan,[1,2,3] and Ariando[1,2,a)]

[1]*NUSNNI-NanoCore, National University of Singapore, 117411 Singapore*

[2]*Department of Physics, National University of Singapore, 117542 Singapore*

[3]*Department of Electrical and Computer Engineering, National University of Singapore, 117576 Singapore*

[4]*CNR-SPIN, Complesso MonteSantangelo via Cinthia, 80126 Napoli, Italy*

[5]*School of Physics and CRANN, Trinity College, Dublin 2, Ireland*

a)e-mail: ariando@nus.edu.sg

**Localization of electrons in the two-dimensional electron gas at the LaAlO$_3$/SrTiO$_3$ interface is investigated by varying the channel thickness in order to establish the nature of the conducting channel. Layers of SrTiO$_3$ were grown on NdGaO$_3$ (110) substrates and capped with LaAlO$_3$. When the SrTiO$_3$ thickness is ≤ 6 unit cells, most electrons at the interface are localized, but when the number of SrTiO$_3$ layers is 8-16, the free carrier density approaches $3.3\times10^{14}$ cm$^{-2}$, the value corresponding to charge transfer of 0.5 electron per unit cell at the interface. The number of delocalized electrons decreases again when the SrTiO$_3$ thickness is ≥ 20 unit cells. The ~4 nm conducting channel is therefore located significantly below the interface. The results are explained in terms of Anderson localization and the position of the mobility edge with respect to the Fermi level.**

The two dimensional electron gas (2DEG) at the interface between the band insulators $LaAlO_3$ and $SrTiO_3$ [1] continues to stimulate the interest of condensed matter researchers. It exhibits a variety of unexpected properties such as superconductivity [2], magnetism [3], and electronic phase separation [5-9]. Recent observation of high mobility at low temperatures (> 5X10$^4$ cm$^2$V$^{-1}$s$^{-1}$) [10] and fabrication of millions of transistors on a single chip [11] have highlighted the importance of these oxide interfaces both from fundamental and applied perspectives. The 2DEG is thought to result from an electron transfer to the interface between the polar oxide ($LaAlO_3$) and the nonpolar oxide ($SrTiO_3$), which is necessary to avoid a divergence of the energy associated with the electric field [12]. A charge transfer of 0.5 electron per interface unit cell (uc) or $3.3\times10^{14}$ cm$^{-2}$ should be required to compensate the electric field in polar $LaAlO_3$ and avert the polar catastrophe [12]. However, a major puzzle is that the experimentally-observed carrier densities at low temperatures for the 2DEG in fully oxidized samples [3,5,13] are an order of magnitude lower than expected. Furthermore, it is unclear where exactly at the interface the conduction electrons are located, since the $LaAlO_3$ is usually grown on a $SrTiO_3$ substrate. One proposed explanation is that these 'disappearing' electrons are localized within the first $SrTiO_3$ layers that are closest to the interface, where Ti $3d_{xy}$ sub-bands have lower energy than the other Ti $3d$ orbitals [14-16]. According to the theoretical calculations, the mobile electrons responsible for the transport properties of the 2DEG are farther away (≥ 4 uc) from the interface [15]. While experimental results [12,17-19] have shown that the 2DEG can penetrate some distance from the interface into $SrTiO_3$ layer, the exact location of the conduction electrons has not been determined. Since the conventional way to fabricate $LaAlO_3$/$SrTiO_3$ interfaces is to grow $LaAlO_3$ layers on $SrTiO_3$ substrates, it had not been possible to determine the location of the conducting channel.

A drawback with the conventional LaAlO$_3$/SrTiO$_3$ interfaces [1-13] is that losses in the SrTiO$_3$ substrate limit the potential application of the 2DEG in high frequency devices [20]. In order to overcome this limitation and expand the applicability of 2DEG, LaAlO$_3$/SrTiO$_3$ interfaces have been fabricated on other substrates such as silicon [21], NdGaO$_3$ (110) [22], (LaAlO$_3$)$_{0.3}$(Sr$_2$AlTaO$_6$)$_{0.7}$ (001) [22-24], and DyScO$_3$ (110) [22]. Such an approach may permit a demonstration of the novel *topological superconductivity* recently predicted to appear in two-layer interacting Rashba systems, which might be fabricated by growing LaAlO$_3$/SrTiO$_3$ interfaces on LaAlO$_3$ substrates [25]. The main aim of the present work is to understand how the physical properties of the 2DEG vary with thickness of the SrTiO$_3$ layer at the interface, and to study the transport properties of the electrons involved.

We have grown the LaAlO$_3$/SrTiO$_3$ heterostructure on NdGaO$_3$ (110) substrates. The crystal structure of NdGaO$_3$ is indexed on an orthorhombic (√2$a_0$, √2$a_0$, 2$a_0$) type cell with $a$ = 5.433 Å, $b$ = 5.503 Å and $c$ = 7.716 Å. When indexed on the pseudocubic cell ($a_0$, $a_0$, $a_0$), the in-plane lattice constants $a_0$ for the pseudocubic lattices of SrTiO$_3$ (100), NdGaO$_3$ (110), and LaAlO$_3$ (100) are 3.905, 3.858, and 3.790 Å, respectively. It is therefore possible to grow epitaxial LaAlO$_3$/SrTiO$_3$ (100) interfaces on NdGaO$_3$ (110) substrates. Ideally, a LaAlO$_3$ (100) substrate would be the best candidate due to its capacity to reduce the lattice-mismatch-induced strain at the interface, and its low loss tangent at high frequencies [20,26]. However, problems of crystal twinning [27] and unstable surface termination [28] in LaAlO$_3$ set a very stringent limit, as for using a LaAlO$_3$/SrTiO$_3$/LaAlO$_3$ structure for the observation of topological superconductivity [25], and this leads us to choose NdGaO$_3$ (110) instead. At the conventional LaAlO$_3$/SrTiO$_3$ interface, the lattice mismatch between LaAlO$_3$ and SrTiO$_3$ is 3%, and it is responsible for the interface strain. However, when the SrTiO$_3$ layer is grown on NdGaO$_3$ (110), the large mismatch between LaAlO$_3$ and SrTiO$_3$ will be partially transferred from the LaAlO$_3$/SrTiO$_3$ interface to the SrTiO$_3$/NdGaO$_3$ interface. To investigate the depth dependence of the 2DEG, we varied the SrTiO$_3$ thickness *t* from 3 to 25 uc during the fabrication of LaAlO$_3$/SrTiO$_3$/NdGaO$_3$ heterostructures, which were capped with 15 uc of LaAlO$_3$ (Fig. S1 in Supplementary Materials). Before deposition, the NdGaO$_3$ substrates were annealed at 1050 $^{O}$C in air for 2.5 hours to obtain the atomically-flat B-site terminated surfaces [29]. The growth parameters for both SrTiO$_3$ and LaAlO$_3$ layers are as follows: 1.8 J/cm$^2$ for laser energy, 760 $^{O}$C for temperature, and 2×10$^{-4}$ Torr for oxygen partial pressure during the deposition.

The results on temperature-dependent sheet resistance ($R_S$-$T$) are shown in Fig. 1(a). Although at high temperatures the resistance of all the LaAlO$_3$/SrTiO$_3$/NdGaO$_3$ samples is dominated by electron-electron scattering with an $R_S \propto T^2$, an upturn of sheet resistance invariably occurs below a temperature $T_{min}$ (where $R_S$ is minimum). This upturn in $R_S$-$T$ has been also reported at the LaAlO$_3$/SrTiO$_3$ interfaces grown on other substrates [21,23,24], and it depends on the stoichiometry of the LaAlO$_3$ [30]. Reducing the thickness of the SrTiO$_3$ layer raises $T_{min}$ and $R_S$ at the same time. For the samples with 3, 4, and 6 uc SrTiO$_3$ layers, the $R_S$ at 2 K is far above the quantum of resistance (12.9 kΩ, including spin degeneracy). The $R_S$-$T$ curves for these samples diverge as the temperature is decreased, and they can be well fitted to $R_S \propto \exp[(T_0/T)^{1/2}]$, which suggests carrier localization in these heterostructures and modified variable range hoping (VRH) with a soft two-dimensional Coulomb gap [31]. For comparison, the samples with the thicker SrTiO$_3$ (8 and 12 uc) have $R_S$ less than half of 12.9 kΩ, and the $R_S$-$T$ curves are of the form $1/R_S \propto A + B\ln T$ (A and B are constants) below $T_{min}$ suggesting weak localization in two-dimensions [32]. Given our experimental result that the SrTiO$_3$/NdGaO$_3$ heterostructures prepared in 10$^{-4}$ Torr are insulating ($R_S > 10^7$ Ω), the conducting behavior ($dR_S/dT > 0$ and only weak localization at low temperatures) seen in Fig. 1(a) must be due to the electrons at the LaAlO$_3$/SrTiO$_3$ interface.

This insulating SrTiO$_3$/NdGaO$_3$ interface is different from the conducting one, found when NdGaO$_3$ is grown on a SrTiO$_3$ substrate [33]. In the former case, there is no observable conductivity, similar to the insulating interface, where the SrTiO$_3$ layers are grown on a LaAlO$_3$ substrate [34]. This can be ascribed to the loss of polar discontinuity at the interface. When the NdGaO$_3$ or LaAlO$_3$ is not a freshly-deposited polar layer, but the substrate itself which has been exposed to the ambient atmosphere, the surface charge is compensated by some external charge centers, as the surface of NdGaO$_3$ or LaAlO$_3$ substrate becomes neutral. Hence, the polar discontinuity cannot be easily established at the interface when polar oxides are used as substrates. Similar results to those in Fig. 1(a) are observed when the NdGaO$_3$ (110) substrates are replaced by (LaAlO$_3$)$_{0.3}$(Sr$_2$AlTaO$_6$)$_{0.7}$ (001), which shows that bandgap mismatch with the substrate is not a critical factor.

Moreover, the conducting behavior of the LaAlO$_3$/SrTiO$_3$/NdGaO$_3$ samples also depend greatly on the LaAlO$_3$ thickness, as shown in Fig. 1(b), where it is seen that 10-12 uc of LaAlO$_3$ are needed to make 12 uc of SrTiO$_3$ conducting. But this critical thickness of LaAlO$_3$ is larger than 4 uc that is commonly observed at the conventional LaAlO$_3$/SrTiO$_3$ interfaces. Given that a higher critical thickness for the LaAlO$_3$/SrTiO$_3$ interfaces grown on other substrates is also observed by Bark *et al.*, who found that with 50 uc SrTiO$_3$ the critical LaAlO$_3$ thickness is around 15 uc [22], this result could imply an important role of the strain in this phenomenon. Figure 1(c) shows the variation of $T_{min}$ with SrTiO$_3$ thickness $t$. The temperature $T_{min}$ separates the regions with $dR_S/dT > 0$ for the higher temperatures and thicker SrTiO$_3$ layers, and $dR_S/dT < 0$ on the opposite side. The dependence of $T_{min}$ on SrTiO$_3$ thickness can be described by $T_{min} \sim 1/t$ [Fig. S6(c)], which is consistent with a temperature-dependent mean free path or relaxation time signifying small energy transfer scattering in the 2DEG [35-37].

The temperature dependence of sheet carrier density ($n_S$) and mobility ($\mu_H$) are plotted in Fig. 2(a) and 2(b) respectively. Compared to conventional LaAlO$_3$/SrTiO$_3$ interfaces grown on SrTiO$_3$ substrates [3,5], the interfaces grown epitaxially on NdGaO$_3$ with an 8-16 uc SrTiO$_3$ layer show an $n_S$ independent of temperature below 100 K that is one order of magnitude larger, and close to the value of 3.3×10$^{14}$ cm$^{-2}$, which corresponds to 0.5 electrons per unit cell at the interface . It can be argued that the reason for the temperature independence is the clamping effect of the NdGaO$_3$ substrate, which prevents the SrTiO$_3$ layer from undergoing the structural transitions that occur at low temperatures [5,38], thus avoiding the strong localization of carriers at low temperatures, which is widely observed in the 2DEG on SrTiO$_3$ substrates. This result supports the view that the SrTiO$_3$ phase transitions are important for determining the low temperature 2DEG properties, carrier localization in particular [39]. Also the absence of temperature dependence in mobility when the free carrier concentration is high suggests that electron-electron scattering is dominant. The main point here is that these characteristics, the temperature-independent value of $n_S$ and high density of free carriers approaching the ideal value predicted by the polar catastrophe model, are observed only when the number of SrTiO$_3$ monolayers is ≥ 8. This implies that the formation of a mobile 2DEG requires at least 8 uc (~ 3 nm) of SrTiO$_3$ for the conducting channel, which should be regarded as the minimal propagating depth for the 2DEG and is consistent with previous experiments performed on conventional LaAlO$_3$/SrTiO$_3$ heterostructures [12,17], but this is the first time that the opposite limit, i.e. carrier localization induced in SrTiO$_3$ layers ≤ 6 uc thick, has been directly observed.

In order to illustrate further the effect of SrTiO$_3$ thickness on the 2DEG, room-temperature $n_S$ and $\mu_H$ as a function of SrTiO$_3$ thickness $t$ are plotted in Fig. 3(a) and 3(b). The abrupt enhancement of $n_S$ from 0.9×10$^{14}$ to 2.9×10$^{14}$ cm$^{-2}$ is observed in a very small window of SrTiO$_3$ thicknesses, i.e. from 6 to 8 uc. The value of $n_S$ seems to saturate for $t$ = 8-16 uc, and then falls to 0.8×10$^{14}$ and 0.6×10$^{14}$ cm$^{-2}$ at 20 and 25 uc. This is also consistent with the results of Bark et al., who observed a low value of $n_S$ = 0.5×10$^{14}$ cm$^{-2}$ at 50 uc [22]. On the other hand, in Fig. 3(b), a linear increase of room-temperature $\mu_H$ with SrTiO$_3$ thickness is observed from 3 to 12 uc, which proves that the SrTiO$_3$ layer is truly the conducting channel for the 2DEG. On further increasing the SrTiO$_3$ thickness to 16 and 25 uc, a nearly-constant $\mu_H$ (close to that for the conventional LaAlO$_3$/SrTiO$_3$ interface at room temperature) is attained. According to the above data, the SrTiO$_3$ thickness required for a mobile 2DEG is around 8-16 uc.

A rough estimate of the width $t$ of the 2DEG at the interface can be obtained by considering its energy per unit area [Fig. 3(c) and Fig. S9]. Generally, the electrons can lower their energy by spreading out deeper into the SrTiO$_3$, provided the states are available. However, there is an energy penalty to be paid because the polarization fields extend into the SrTiO$_3$, hence the total energy to confine the 2DEG in SrTiO$_3$ can be written as:

$$E_{2DEG} \approx n\hbar^2/2mt^2 + \sigma^2 t/24\varepsilon_0\varepsilon, \tag{1}$$

where n is the electron density per unit area, m is the electron mass, σ is the sheet charge density and ε is the dielectric constant of SrTiO$_3$. Minimizing this energy with respect to $t$, we find the minimum at

$$t = [24\varepsilon_0\varepsilon n\hbar^2/m\sigma^2]^{1/3}, \tag{2}$$

with n = 3.3×10$^{14}$ cm$^{-2}$, σ = 50 μC cm$^{-2}$ and ε = 300, a value that depends little on temperature in SrTiO$_3$ thin films [40], this gives us a SrTiO$_3$ thickness of 2.2 nm or 6 uc, which is close to our minimal SrTiO$_3$ thickness of 8 uc for a delocalized 2DEG.

We propose that the electron transport is dominated by Anderson localization, which is related to the two-dimensional nature of the channel. The delocalized carrier density depends on the effective position of the mobility edge ($E_M$) with respect to the Fermi level ($E_F$). When the SrTiO$_3$ layer is very thin, the electrons are localized because of the potential fluctuations induced by charge disorder due to ionic interdiffusion at the interface. This ionic interdiffusion usually involves the first 1-2 uc from the interface, and it will create a mobility edge, localizing the states at the bottom of the band. As the SrTiO$_3$ thickness increases, these potential fluctuations are screened, and the mobility edge drops rapidly to a lower energy. The Fermi energy falls less rapidly with thickness, varying as as $1/t$ for a constant density of states. At about 8 uc of SrTiO$_3$ $E_F$ exceeds $E_M$ and we see the onset of electron delocalization, and metallic conductivity at low temperatures. Our experiments clearly prove that the conducting electrons are located appreciably below the LaAlO$_3$/SrTiO$_3$ interface of our films. The reduction in carrier density again in SrTiO$_3$ layers thicker than 16 uc could be understood in several ways. As the 0.5 electrons per Ti extend into SrTiO$_3$ layer further from the interface, the Fermi level $E_F$ falls towards the bottom of the band due to the increase of available electronic states and it approaches the $t_{2g}$ band mobility edge for the bulk. This arises from defects such as oxygen vacancies that are distributed throughout the SrTiO$_3$, which is more disordered when it is an epitaxially-grown layer rather than a single-crystal substrate. Also, there is a tendency for strain relaxation and associated defect production in these thicker SrTiO$_3$ layers, which can raise the mobility edge again and localize the carriers. A third possibility is the polarization of the distorted SrTiO$_3$ layer, which can compensate the polarization catastrophe at the interface [22].

We can model the number of mobile electrons with thickness quite nicely with a single band having a constant density of states. The total number of available states increases in proportion to the number of layers, leading to a Fermi level (relative to the bottom of the rectangular conduction band)

$$E_F = a/t, \qquad (3)$$

where $t$ is the SrTiO$_3$ thickness and $a$ is a constant. Then we model the mobility edge to vary due to two effects - one falling off exponentially with $t$, due to the disorder in the interface layer; the other a constant low-energy mobility edge due to residual disorder in the SrTiO$_3$, giving

$$E_M = b\exp(-t/t_0) + c, \qquad (4)$$

where $b$, $t_0$ and $c$ are all constants. At room temperature, we roughly assume that there is a baseline for the delocalized carrier density $n_S$ due to the thermal activation. Hence, $n_S$ can be evaluated by

$$n_S(E_F) = \begin{cases} d & (E_F < E_M) \\ d + n\left[\frac{E_F - E_M}{E_F}\right] & (E_F \geq E_F) \end{cases} \qquad (5)$$

where $d$ is a constant baseline (assumed to be $0.5\times10^{14}$ cm$^{-2}$ which is the carrier density for 50 uc SrTiO$_3$ [22]) and $n$ is the carrier density for the ideal 2DEG, $3.3\times10^{14}$ cm$^{-2}$. Based on these, we show the fit to both $E_F$ and $E_M$ as a function of thickness in Fig. 4(a). Because $E_F$ falls much more slowly with thickness and $E_M$ is higher than $E_F$ at low thickness, there are two crossover points of $t_1$ and $t_2$, indicating free carriers in the thickness range $t_1 < t < t_2$. The fitted curve for the delocalized carriers is also quite consistent with the experimentally observed values in Fig. 4 (b). A reasonable screening length $t_0$ of about 1-2 unit cells is seen for the mobility edge, details are of which given in Fig. S10.

In conclusion, we have shown that the thickness of the SrTiO$_3$ layer in epitaxially-grown LaAlO$_3$/SrTiO$_3$ heterostructures is critical for determining the 2DEG transport properties. Samples with the thinnest SrTiO$_3$ exhibit a low carrier density at room temperature, a robust insulating ground state, and modified variable-range hopping transport behavior at low temperatures, due to the Anderson localization of the 2DEG arising from the random interface potential with ionic interdiffusion. When the SrTiO$_3$ thickness reaches 8-16 uc, the carrier density increases to almost the expected 0.5 electrons per interface Ti site, and the mobility saturates at the conventional room-temperature value. However, the carrier density falls again for the thicker layers to the conventional value found for single-crystal SrTiO$_3$ substrates. Most of the Ti 3*d* electrons are localized at the very bottom of the 3*d* band, below the mobility edge. We are therefore able to explain the observed behavior on a localization model where the position of the Fermi level and the mobility edge depend on the SrTiO$_3$ layer thickness. The study of these thin epitaxially-grown layers has enabled us to describe the role of localization and to define the extent of the conducting region at the LaAlO$_3$/SrTiO$_3$ interface. It shows how to tailor the oxide interface to optimize the two-dimensional conduction, which will be of importance for oxide electronics.

## Acknowledgments

This work was financially supported by the National Research Foundation (NRF) Singapore under the Competitive Research Program 'Tailoring Oxide Electronics by Atomic Control: Oxides with Novel Functionality' [R-398-000-062-281], NUS cross-faculty grant and FRC.

1. A. Ohtomo and H. Y. Hwang, Nature **427**, 423 (2004).

2. N. Reyren *et al.*, Science **317**, 1196 (2007).

3. A. Brinkman *et al.*, Nature Mater. **6**, 493 (2007).

4. A. D. Caviglia *et al.*, Phys. Rev. Lett. **104**, 126803 (2010); M. Ben Shalom *et al.*, *ibid*. **104**, 126802 (2010).

5. Ariando *et al.*, Nature Commun. **2**, 188 (2011).

6. D. A. Dikin *et al.*, Phys. Rev. Lett. **107**, 056802 (2011).

7. J. A. Bert *et al.*, Nature Phys. **7**, 767 (2011).

8. B. Kalisky *et al.*, Nature Commun. **3**, 922 (2012).

9. L. Li, C. Richter, J. Mannhart, and R. C. Ashoori, Nature Phys. **7**, 762 (2011).

10. M. Huijben *et al.*, Adv. Funct. Mater. doi: 10.1002/adfm.201203355.

11. B. Förg, C. Richter, and J. Mannhart, Appl. Phys. Lett. **100**, 053506 (2012).

12. N. Nakagawa, H. Y. Hwang, and D. A. Muller, Nature Mater. **5**, 204 (2006).

13. C. Cancellieri *et al.*, Europhys. Lett. **91**, 17004 (2010).

14. Z. S. Popović, S. Satpathy, and R. M. Martin, Phys. Rev. Lett. **101**, 256801 (2008).

15. P. Delugas *et al.*, Phys. Rev. Lett. **106**, 166807 (2011).

16. M. Salluzzo *et al.*, Phys. Rev. Lett. **102**, 166804 (2009).

17. M. Basletic *et al.*, Nature Mater. **7**, 621 (2008).

18. O. Copie *et al.*, Phys. Rev. Lett. **102**, 216804 (2009).

19. T. Fix, J. L. MacManus-Driscoll, and M. G. Blamire, Appl. Phys. Lett. **94**, 172101 (2009).

20. J. M. Phillips, J. Appl. Phys. **79**, 1829 (1996).

21. J. W. Park *et al.*, Nature Commun. **1**, 94 (2010).

22. C. W. Bark *et al.*, P. Natl. Acad. Sci. USA. **108**, 4720 (2011).

23. P. Brinks *et al.*, Appl. Phys. Lett. **98**, 242904 (2011).

24. T. Hernandez *et al.*, Phys. Rev. B **85**, 161407 (2012).

25. S. Nakosai, Y. Tanaka, and N. Nagaosa, Phys. Rev. Lett. **108**, 147003 (2012).

26. C. Zuccaro, M. Winter, N. Klein, and K. Urban, J. Appl. Phys. **82**, 5695 (1997).

27. Z. L. Wang and A. J. Shapiro, Surf. Sci. **328**, 141 (1995).

28. J. Yao *et al.*, J. Chem. Phys. **108**, 1645 (1998).

29. R. Gunnarsson, A. S. Kalabukhov, and D. Winkler, Surf. Sci. **603**, 151 (2009).

30. E. Breckenfield *et al*, Phys. Rev. Lett. **110**, 196804 (2013).

31. V. Yu. Butko, J. F. DiTusa, and P. W. Adams, Phys. Rev. Lett. **84**, 1543 (2000).

32. P. A. Lee and T. V. Ramakrishnan, Rev. Mod. Phys. **57**, 287 (1985); A. D. Caviglia *et al.*, Nature **456**, 624 (2008).

33. A. Annadi *et al.*, Phys. Rev. B **86**, 085450 (2012); U. S. di. Uccio *et al.*, arXiv:1206.5083.

34. Z. Q. Liu *et al.*, AIP Advances **2**, 012147 (2012).

35. B. L. Altshuler, A. G. Aronov, and D. E. Khmelnitsky, J. Phys. C: Solid State Phys. **15**, 7367 (1982).

36. B. N. Narozhny, G. Zala, and I. L. Aleiner, Phys. Rev. B **65**, 180202 (2002).

37. I. R. Pagnossin, A. K. Meikap, A. A. Quivy, and G. M. Gusev, J. Appl. Phys. **104**, 073723 (2008).

38. F. W. Lytle, J. Appl. Phys. **35**, 2212 (1964).

39. W. M. Lü *et al.*, Appl. Phys. Lett. **99**, 172103 (2011).

40. A. Walkenhorst *et al.*, Appl. Phys. Lett. **60**, 1744 (1992).

Fig. 1. (a) $R_S$-$T$ curves for LaAlO$_3$/SrTiO$_3$/NdGaO$_3$ heterostructures with fixed LaAlO$_3$ thickness (15 uc) and different SrTiO$_3$ thickness (from 3 to 12 uc). The blue, dark cyan and black lines are fits using electron-electron scattering, modified VRH with a two-dimensional Coulomb gap and weak localization models, respectively. The arrows indicate the upturn temperatures, $T_{min}$. (b) Room-temperature sheet conductivity as a function of LaAlO$_3$ thickness for a fixed SrTiO$_3$ thickness of 12 uc, and also as a function of SrTiO$_3$ thickness for a fixed LaAlO$_3$ thickness of 15 uc. The red dash line is the measurement limitation. Inset is the schematic view for layer structures in LaAlO$_3$/SrTiO$_3$/NdGaO$_3$ [LAO(100)/STO(100)/NGO(110)] heterostructure, in which the 2DEG exists at the LaAlO$_3$/SrTiO$_3$ interface. (c) 'Metal-insulator' phase diagram of LaAlO$_3$/SrTiO$_3$/NdGaO$_3$ heterostructure versus SrTiO$_3$ thickness with a fixed LaAlO$_3$ thickness at 15 uc. The slope of the red dashed line is -1.

Fig. 2. (a) The carrier density $n_S$ and (b) mobility $\mu_H$ as a function of temperature for samples with different SrTiO$_3$ thickness, from 3 to 12 uc keeping the LaAlO$_3$ thickness at 15 uc.

Fig. 3. The SrTiO$_3$-thickness-dependent $n_S$ and $\mu_H$ at 300 K are shown in (a) and (b), respectively. (c) The calculation of 2DEG energy per unit area of interface, and the details are discussed with Fig. S9. (d) Schematic view of the LaAlO$_3$/SrTiO$_3$/NdGaO$_3$ heterostructure showing at low temperatures the SrTiO$_3$ thickness range for a mobile 2DEG is around 8-16 uc (green area), and below 6 uc the carriers are localized (blue area).

Fig. 4. (a) Fitted $E_F$ and $E_M$ as a function of SrTiO$_3$ thickness. (b) The comparison on experimentally observed $n_S$ and fitted $n_S$ as a function of SrTiO$_3$ thickness. The hollow square in (b) is taken from Ref. 20 as comparison. The details on controlling the fitting parameters are discussed in Fig. S10. Schematic 2D-DOS (number of available electronic states per unit interface area) versus energy for $t_{2g}$ band are shown with SrTiO$_3$ thickness $\leq$ 6 uc in (c), 8-16 uc in (d), and $\geq$ 20 uc in (e). Electrons below $E_M$ are localized and denoted by blue, while electrons which are below $E_F$ but beyond $E_M$ are delocalized and denoted by orange. In these sketches, the total area below $E_F$ is fixed for each thickness, indicating the total number of carriers at the LaAlO$_3$/SrTiO$_3$ interface is fixed.

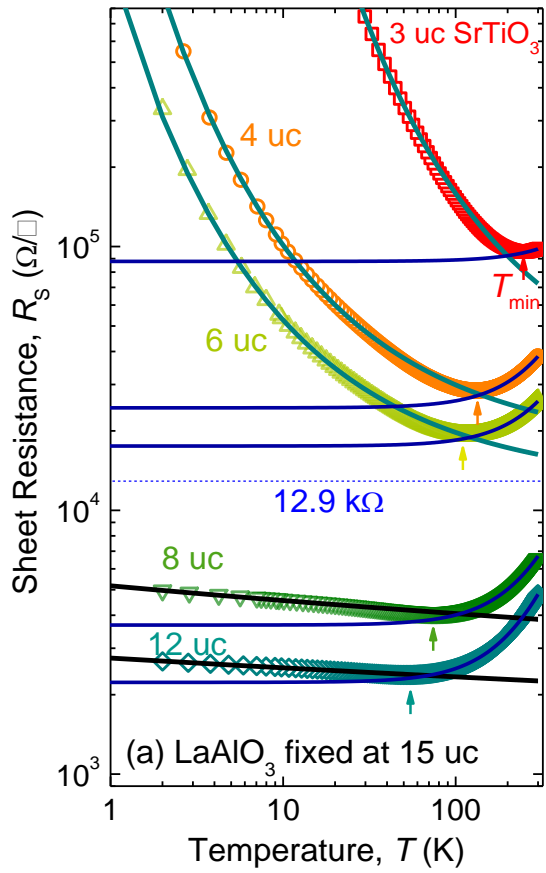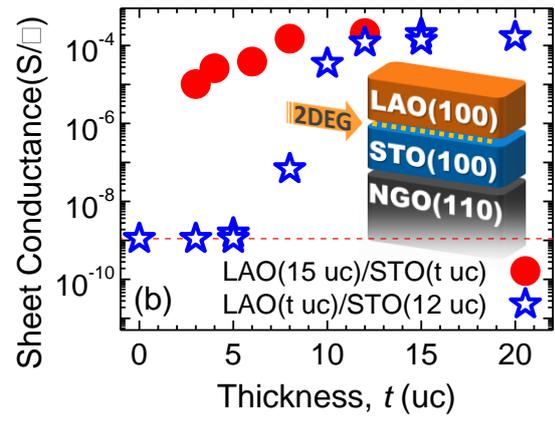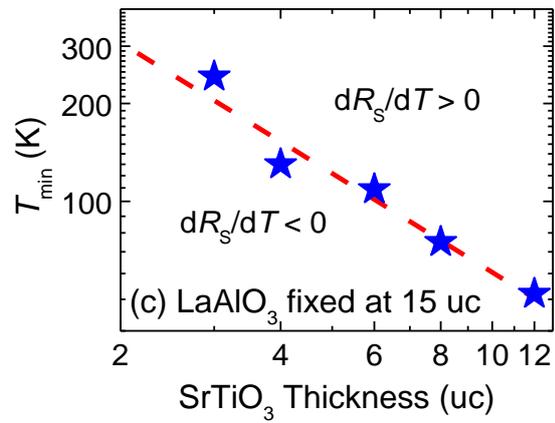

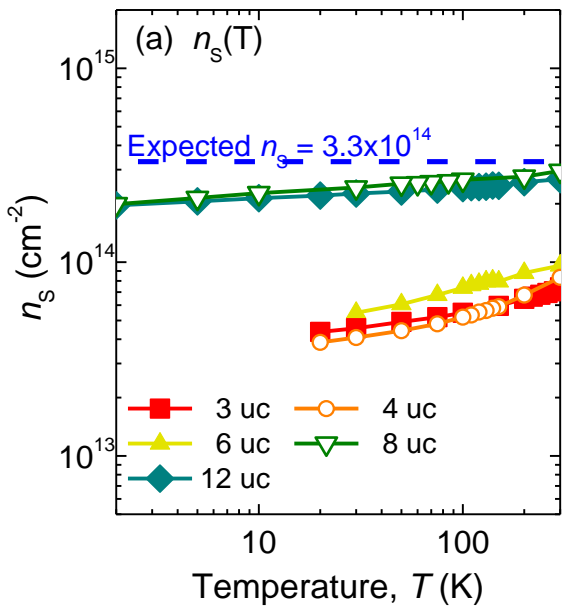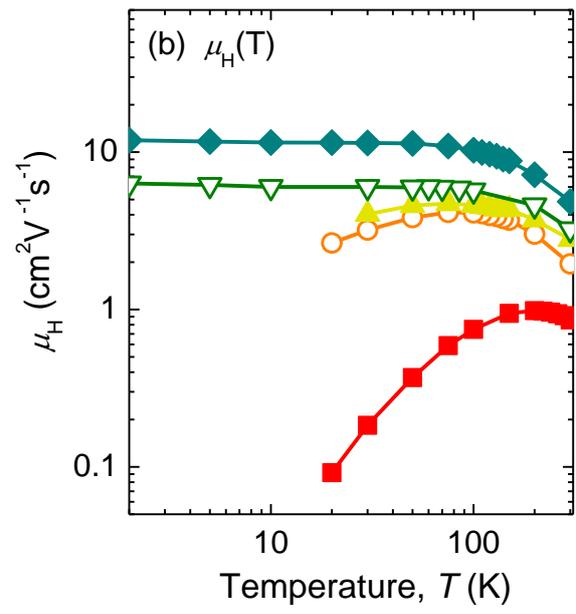

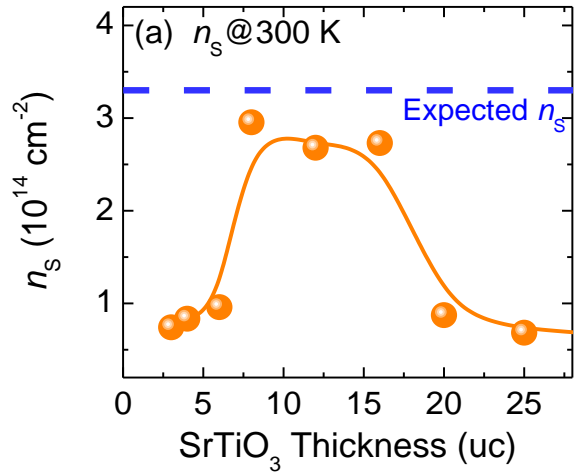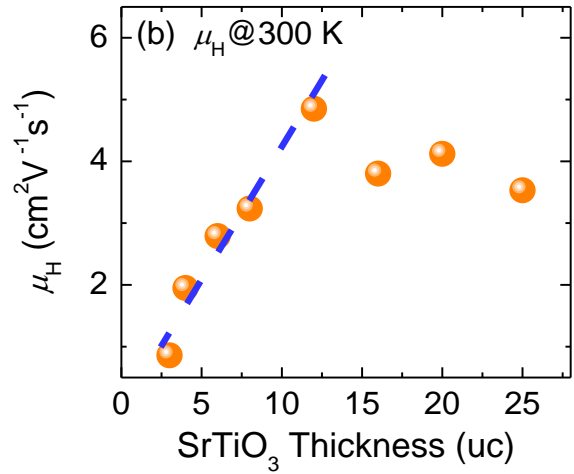
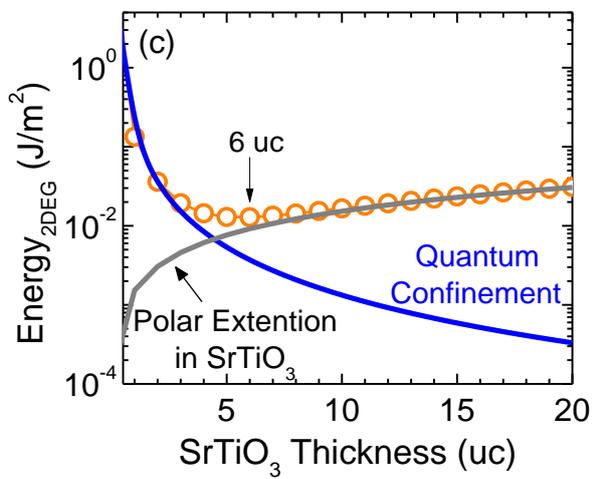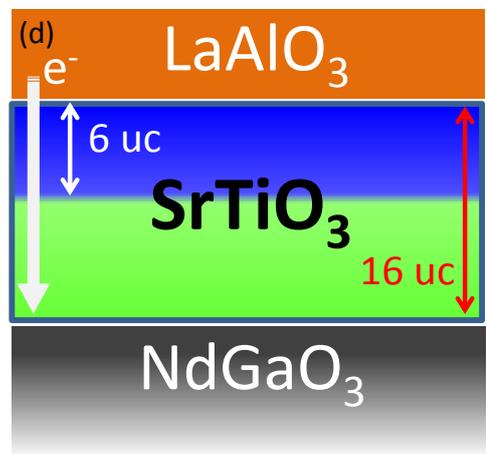

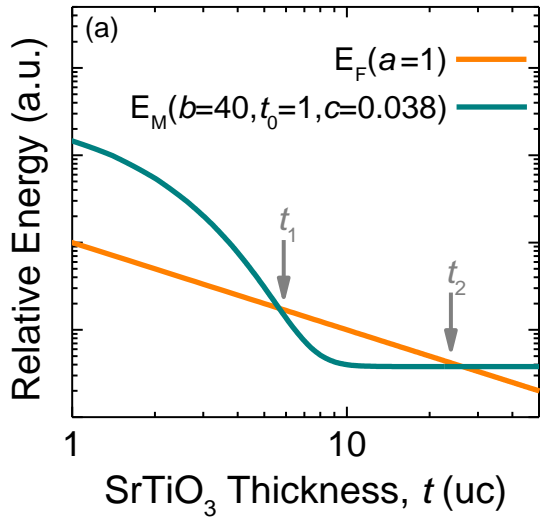
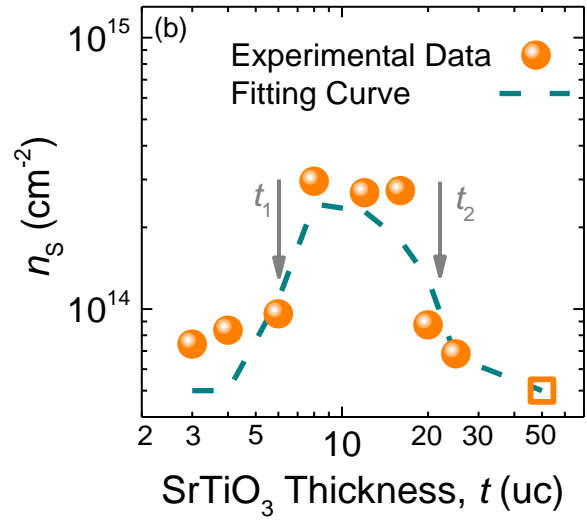
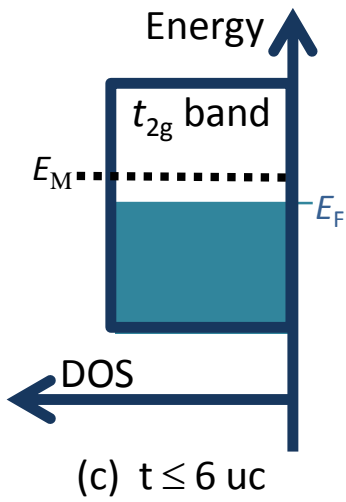
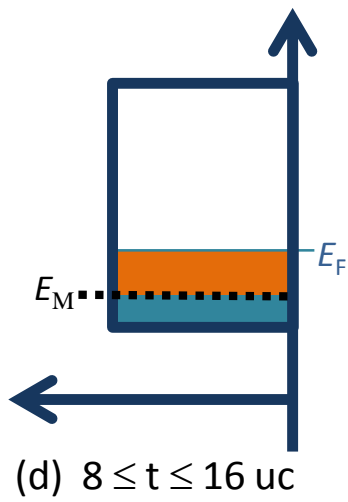
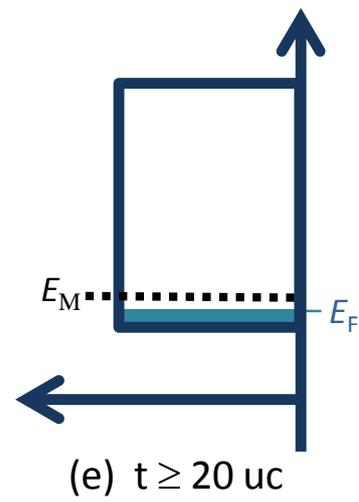

# The Conducting Channel of the LaAlO$_3$/SrTiO$_3$ Interface

## *Supplementary Materials*

Z. Huang,[1] X. Renshaw Wang,[1,2] Z. Q. Liu,[1,2] W. M. Lü,[1,3] S. W. Zeng,[1,2] A. Annadi,[1,2] W. L. Tan,[2] X. P. Qiu,[3] Y. L. Zhao,[1,2] M. Salluzzo,[4] J. M. D. Coey,[1,5] , T. Venkatesan,[1,2,3] and Ariando[1,2,a)]

[1]*NUSNNI-NanoCore, National University of Singapore, 117411 Singapore*
[2]*Department of Physics, National University of Singapore, 117542 Singapore*
[3]*Department of Electrical and Computer Engineering, National University of Singapore, 117576 Singapore*
[4]*CNR-SPIN, Complesso MonteSantangelo via Cinthia, 80126 Napoli, Italy*
[5]*School of Physics and CRANN, Trinity College, Dublin 2, Ireland*

*a)e-mail: ariando@nus.edu.sg*

# I. Interface Preparation

The LaAlO$_3$/SrTiO$_3$/NdGaO$_3$ (LAO/STO/NGO) samples studied here is prepared by depositing LaAlO$_3$ just after depositing SrTiO$_3$ on NdGaO$_3$ (110) substrates. These samples are called the *in-situ grown* samples. The growth parameters for both SrTiO$_3$ and LaAlO$_3$ layers are as follows: 1.8 J/cm$^2$ for laser energy, 760 $^{\circ}$C for temperature, and 2×10$^{-4}$ Torr for oxygen partial pressure during the deposition. In Fig. **S1**, we show the Reflection High-Energy Electron Diffraction (RHEED) oscillations and Atomic Force Microscopy (AFM) images, which indicate the high quality of our samples.

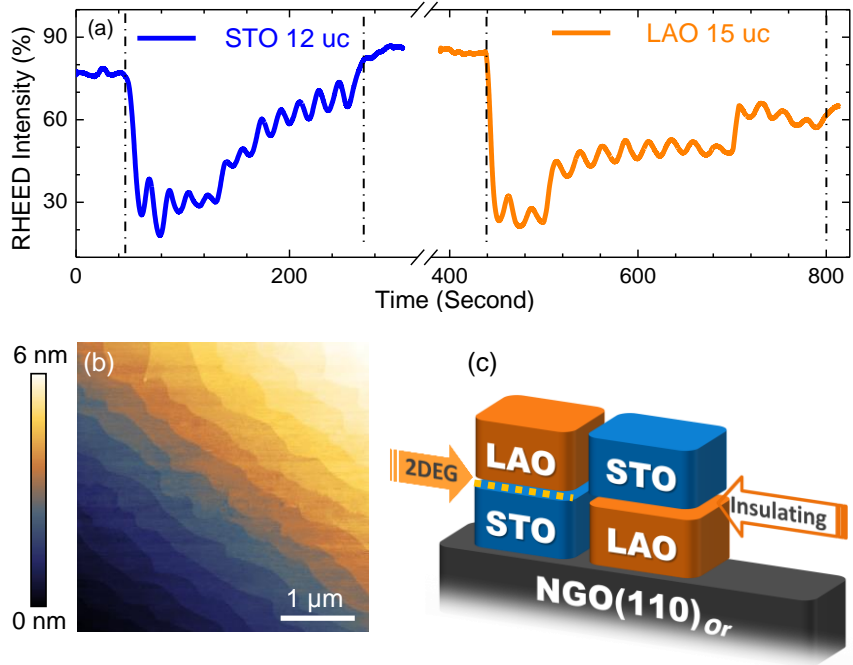

**Figure S1**  (a) RHEED oscillations during deposition for LaAlO$_3$/SrTiO$_3$/NdGaO$_3$ interface. (b) AFM images for the interface with 15 uc LaAlO$_3$ and 12 uc SrTiO$_3$. (c) With the optimized thickness for LAO and STO layers, the interfaces with LaAlO$_3$ as top layers are always conducting, but the ones with SrTiO$_3$ as top layers are always insulating ($R_S > 10^7$ Ω). This can be understood by the different type interfaces (n-type or p-type) formed in the samples.

The main question is does those *in-situ grown* LaAlO$_3$/SrTiO$_3$ have a TiO$_2$ terminated SrTiO$_3$ layer? The AFM images of a freshly grown SrTiO$_3$ layer (12 uc SrTiO$_3$ films on NdGaO$_3$ substrate, Fig. **S2(a)** and **S2(d)**) show step flow growth surfaces. Now we have treated this SrTiO$_3$ layer much like a substrate and did HF etching to ensure a TiO$_2$ termination. AFM image in Fig. **S2(b)** also shows high quality SrTiO$_3$ surface after a proper HF treatment. We denote these SrTiO$_3$/NdGaO$_3$ samples as 'good surface' samples. And if we prolong the duration of HF treatment, as shown in Fig. **S2(e)**, the surface turns bad – the atomic steps become less distinct in AFM image and surface reconstruction is observed in RHEED patterns. We call these SrTiO$_3$/NdGaO$_3$ samples with a 'bad surface'. As shown in Fig. **S2(c)** and **S2(f)**, if the crystalline LaAlO$_3$ is grown on 'good surface', the interface is conducting and shows weak-localization-like behavior at low temperatures identical to the *in-situ grown* (non-HF-treated) sample. This suggests the *in-situ grown* samples are also with TiO$_2$-terminated SrTiO$_3$ layer. And if the crystalline LaAlO$_3$ is grown on 'bad surface', it is completely insulating ($R_S > 10^7$ Ω), showing that the transport property of our samples greatly relies on the SrTiO$_3$ surface.

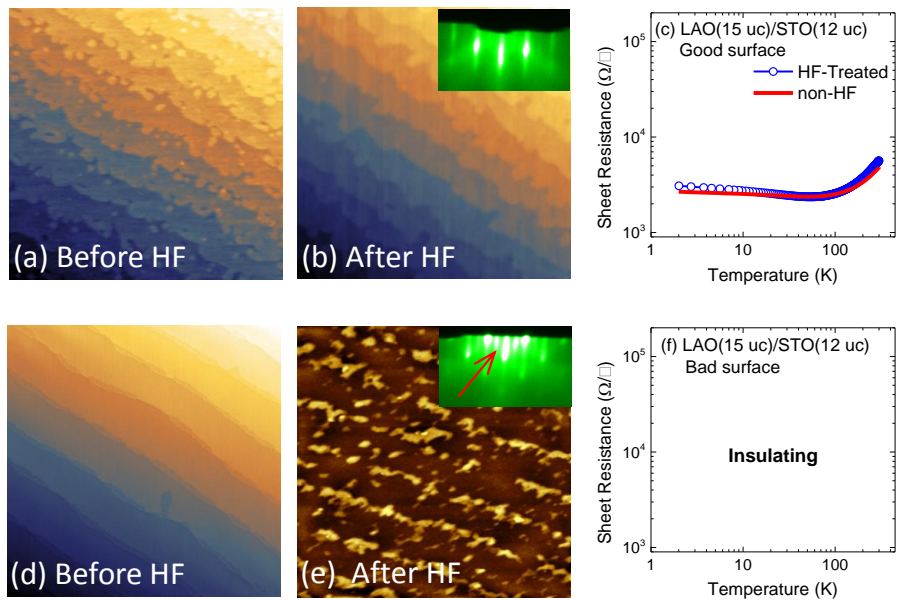

**Figure S2** For 'good surface' sample, AFM images before and after HF treatment are shown in (a) and (b) respectively, RHEED pattern in inset of (b), and $R_S$(T) in (c). There is no obvious difference

between this HF-treated sample and previous non-HF-treated *in-situ grown* sample (also with a good AFM image/surface as shown in Fig. **S1(b)**). For 'bad surface' sample, AFM images before and after HF treatment are shown in (d) and (e) respectively, RHEED pattern in inset of (e), and $R_S$(T) in (f). The additional satellite peak from surface reconstruction is indicated by red arrows.

## II. The Origin of Conductivity

In order to illustrate the effect of polar catastrophe or oxygen vacancy in the conducting LaAlO$_3$/SrTiO$_3$/NdGaO$_3$ samples, we compare below LaAlO$_3$/SrTiO$_3$/NdGaO$_3$ with conventional LaAlO$_3$/SrTiO$_3$ interfaces, and crystalline LaAlO$_3$ on SrTiO$_3$/NdGaO$_3$ with amorphous LaAlO$_3$ on SrTiO$_3$/NdGaO$_3$ under different process conditions.

**a)** For crystalline LaAlO$_3$/(12 uc)SrTiO$_3$/NdGaO$_3$ samples, the critical LaAlO$_3$ thickness is around 10-12 uc. If 10 uc LaAlO$_3$ layers are re-grown on the insulating 5 uc LaAlO$_3$ sample (5+10), the conductivity can be observed and is similar to the conducting 15 uc LaAlO$_3$ sample (orange square in Fig. **S3**). In contrast, if 10 uc LaAlO$_3$ layers are removed from the conducting 15 uc LaAlO$_3$ sample (15-10) by Ar milling, the sample becomes insulating (purple triangle in Fig. **S3**). This Ar-milling experiment supports the fact that the carriers in crystalline LaAlO$_3$/SrTiO$_3$/NdGaO$_3$ arise from polar discontinuity at the LaAlO$_3$/SrTiO$_3$ interface. The similar results on the conventional LaAlO3/SrTiO3 interface can be found at arXiv:1305.5016.

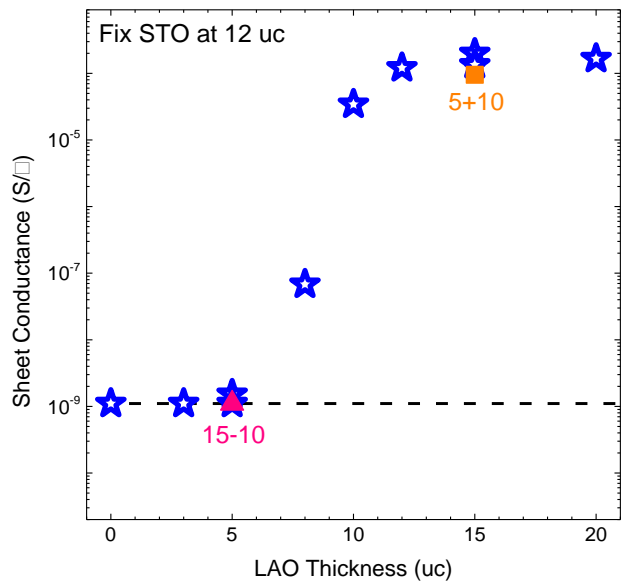

**Figure S3** For crystalline LaAlO$_3$/SrTiO$_3$/NdGaO$_3$ sample with a fixed 12 uc SrTiO$_3$ layer, the sheet conductance changes as a function of LaAlO$_3$ thickness.

**b)** With different duration for HF treatment, we can obtain 'good surface' and 'bad surface' of SrTiO$_3$ layers, as show in Fig. **S2**. For the crystalline LaAlO$_3$/SrTiO$_3$/NdGaO$_3$ sample, the conductivity can only survive when SrTiO$_3$ surface is good. However, when amorphous LaAlO$_3$ is deposited on 'good surface' and 'bad surface' respectively, no obvious difference can be observed as shown in Fig. **S4**. This indicates that the oxygen-vacancy-induced conductivity is not sensitive to the SrTiO$_3$ surface condition in samples with amorphous LaAlO$_3$, and this result is totally different compared to the samples with crystalline LaAlO$_3$. Hence, the origin for conductivity in crystalline LaAlO$_3$/SrTiO$_3$/NdGaO$_3$ has a strong polarization catastrophe component.

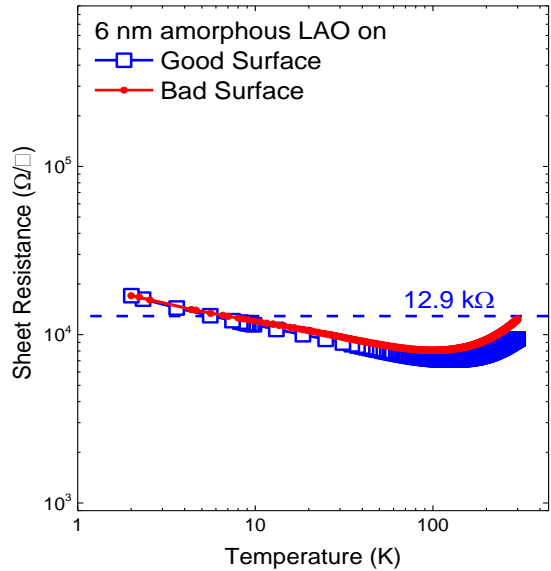

**Figure S4**  The comparison on $R_S(T)$ curves for 6 nm amorphous LaAlO$_3$ layers grown on SrTiO$_3$/NdGaO$_3$ with 'good' and 'bad' surface.

**c)** When changing the SrTiO$_3$ thickness from 12 to 6 uc, the amorphous LaAlO$_3$ samples exhibit different behavior compared to crystalline ones. Especially for amorphous LaAlO$_3$ grown on 6 uc SrTiO$_3$ layers, the sheet resistance at 300 K is below quantum of resistance. But it shows variable-range-hoping $R_S(T)$ behavior at low temperatures.

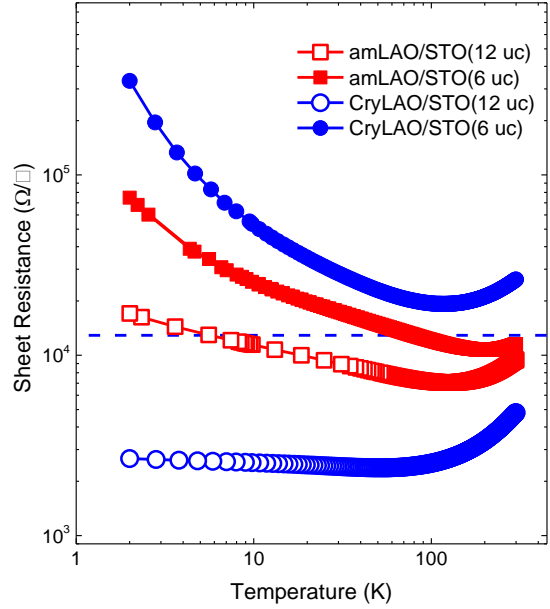

**Figure S5** The comparison on $R_S(T)$ curves for (6 nm amorphous)LaAlO$_3$/SrTiO$_3$/NdGaO$_3$ and (15 uc crystalline)LaAlO$_3$/SrTiO$_3$/NdGaO$_3$, with different SrTiO$_3$ thickness.

Clearly, the results of **a)**-**c)** suggest a definite role for polar discontinuity. However, one strange result that we have not been able to explain is the fact that anneals (ex-situ, 500 °C, in air, 1 h or in-situ cooling, 10 °C/min, 100 mbar) can completely remove the conductivity. This is in contrast to LaAlO$_3$/SrTiO$_3$ where the SrTiO$_3$ is a substrate, in which case the thermal annealing does not get rid of the electrons from the polarization catastrophe. More experiments are needed to clarify the annealing effect on 2DEG at these epitaxially-grown LaAlO$_3$/SrTiO$_3$ interfaces.

# III. Strong Localization and Weak Localization

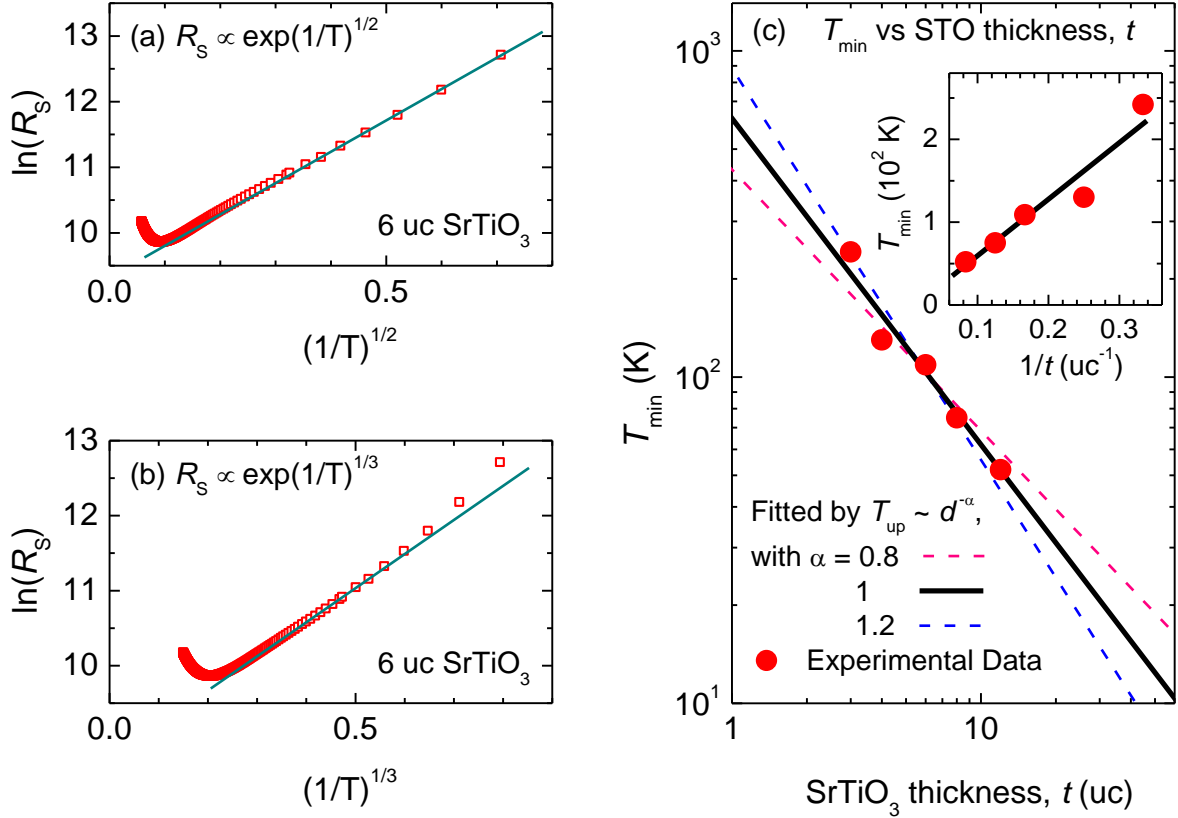

**Figure S6** The fitting of 6 uc $R_S(T)$ by using formula $R_S \propto \exp(1/T)^{1/2}$ in (a) and $R_S \propto \exp(1/T)^{1/3}$ in (b). Clearly, the $R_S(T)$ can be better fitted by the modified variable ranging hopping (VRH) with a two-dimensional Coulomb gap in (a), not the conventional two-dimensional Mott VRH in (b). (c) The fitting of $T_{min}$ as a function of SrTiO$_3$ thickness $t$, where the best-fitted index $\alpha$ is 1. We argue that $R_S(T)$ begins to turn up at low temperature when the mean free path of carrier is very close to the width of conducting channel, or SrTiO$_3$ thickness. Therefore, based on the fitting, the mean free path of carriers ($l$) in our samples should be expressed by $l \propto T^{-1}$, which is consistent with a temperature-dependent mean free path or relaxation time signifying small energy transfer scattering in the 2DEG.

According to our data shown in Fig. 3, when the SrTiO$_3$ thickness increase from 16 to 20(25) uc, the carrier density drops again and $R_S$(300 K) is above the quantum of resistance (12.9 kΩ). Therefore, the strong localization of carriers at low temperatures is expected in those samples, which is proved with the $R_S(T)$ curves shown in Fig. **S7**.

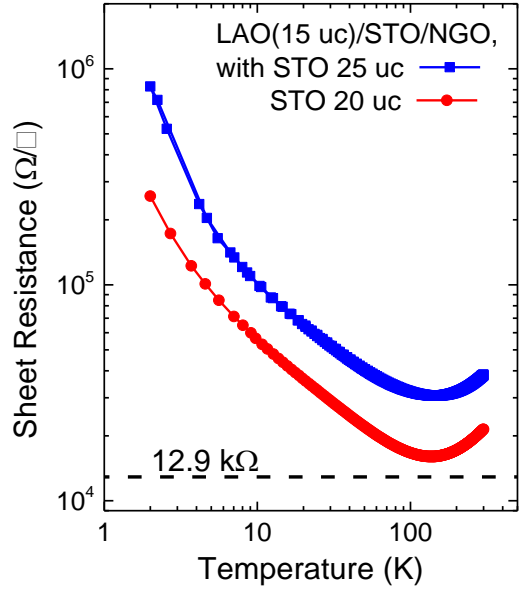

**Figure S7**. $R_S(T)$ curves for LaAlO$_3$(15 uc)/SrTiO$_3$/NdGaO$_3$, when SrTiO$_3$ thickness is 20 and 25 uc.

When SrTiO$_3$ thickness is around 8-16 uc, we argue that the weak localization mechanism dominates at low temperatures. Generally, there are three key signatures of weak localization, the resistance shows divergence and a 1/ln(T) dependence at low temperatures and an anisotropic magneto resistance (MR). While the first two have been shown to be the case in our Fig. 1(a) (in the main text), the MR effect is shown in Fig. **S8**. Although both the in-plane and out-of-plane MR (magnetic field is applied in plane and out of plane when resistance is measured) are positive, the out-of-plane MR still has a negative component when compared with the in-plane one (MR$_{out-of-plane}$ − MR$_{in-plane}$). Therefore, the MR data also suggests weak localization at low temperatures when SrTiO$_3$ thickness is from 8 to 16 uc.

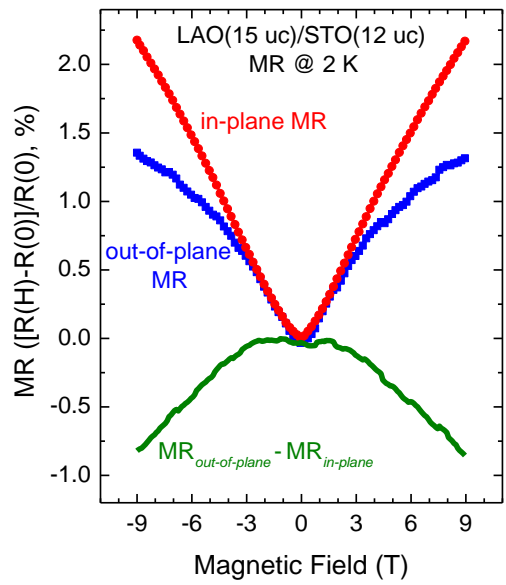

**Figure S8**. In-plane and out-of-plane MR for LaAlO$_3$(15 uc)/SrTiO$_3$(12 uc)/NdGaO$_3$ sample at 2 K

# IV. Calculations on $E_{2DEG}$

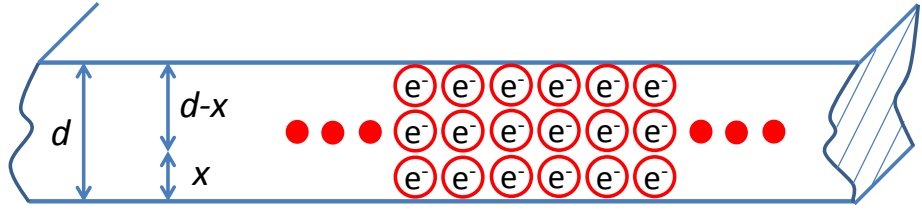

**Figure S9** The sketch for calculation on energy penalty for extending polarization into SrTiO$_3$. The energy of 2DEG with SrTiO$_3$ thickness is combined by two parts. One is from Quantum Confinement ($E_{QC}$), and the other is from Polar Extension in SrTiO$_3$ layers ($E_{PE}$).

Due to the Quantum Effect ($\Delta p * \Delta d \sim \hbar$),

$$E_{QC} \approx n\hbar^2/2md^2 \qquad (1)$$

Where n is the sheet electron density, m is the electron mass, and $d$ is the SrTiO$_3$ thickness.

The energy penalty for polar extension in the charged SrTiO$_3$ ($E_{PE}$) can be calculated as follow.

As shown in above figure, the charged SrTiO$_3$ layers can be thought to be combined by numerous infinite charged sheets. So,

$$Q = \rho V = \sigma S, \qquad \sigma_{sheet} = \sigma/d \qquad (2)$$

Where Q is the total free charge in SrTiO$_3$, ρ is the bulk charge density, V is the volume, σ is the sheet charge density (0.5 Cm$^{-2}$ calculated from 2D carrier density $3.3 \times 10^{18}$ m$^{-2}$), S is the area, and $d$ is the SrTiO$_3$ thickness.

For the layer at position $x$, the electric field $\mathcal{E}(x)$ can be calculated by summing the electric field from all the charged layers in SrTiO$_3$.

Given the formula for electric field of one charged sheet

$$\mathcal{E}_{sheet} = \sigma_{sheet}/2\varepsilon_0\varepsilon \tag{3}$$

Therefore, $\mathcal{E}(x)$ can be written as:

$$\mathcal{E}(x) = \sigma(d-2x)/2\varepsilon_0\varepsilon d \tag{4}$$

Therefore, $E_{PE}$ can be calculated by integral of $\mathcal{E}(x)$.

$$E_{PE} = (1/2)*\varepsilon_0\varepsilon*\int \mathcal{E}(x)dx = \sigma^2 d/24\varepsilon_0\varepsilon \tag{5}$$

Therefore,

$$E_{2DEG} = E_{QC} + E_{PE} \approx n\hbar^2/2md^2 + \sigma^2 d/24\varepsilon_0\varepsilon \tag{6}$$

$$d_{min} = [24\varepsilon_0\varepsilon n\hbar^2/m\sigma^2]^{1/3} \sim 2.2 \text{ nm or 6 uc } SrTiO_3 \tag{7}$$

## V. Fitting of $E_F$ and $E_M$

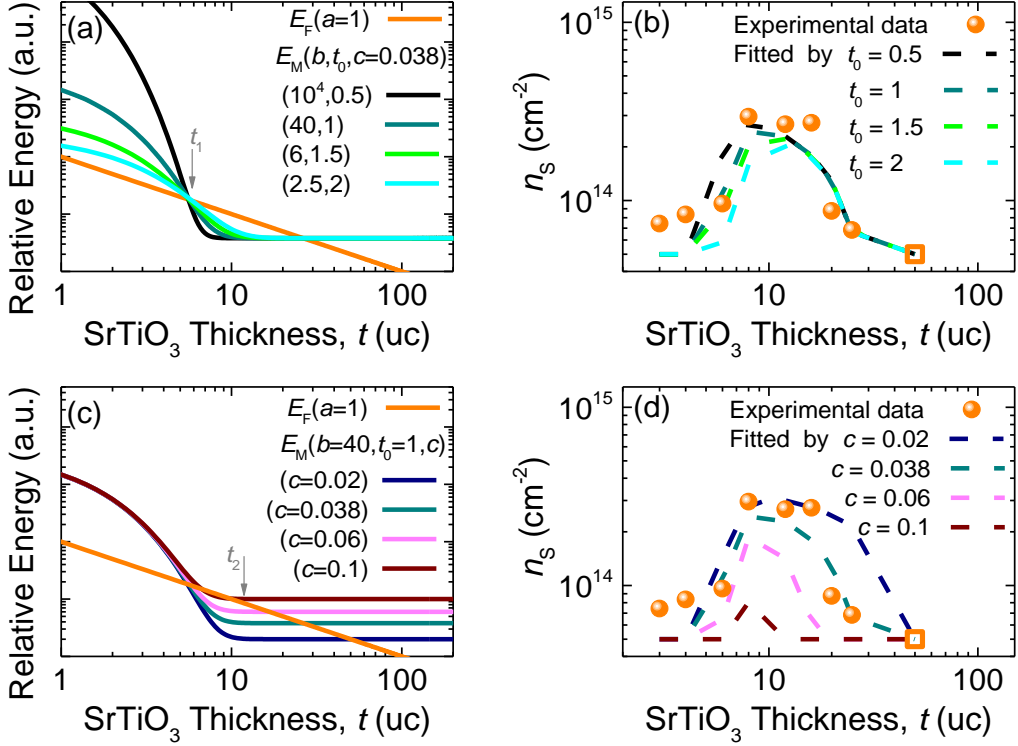

**Figure S10** (a) By fixing $a = 1$ and $c = 0.038$ and changing the value of $b$ and screening length $t_0$, the fitting of $E_F$ and $E_M$ is shown in (a) and $n_S$ in (b). The transition around $t_1$, including the changes of critical thickness and carrier density, is clearly influenced by $b$ and $t_0$. And by fixing $a = 1$, $b = 40$ and $t_0 = 1$ and changing the value of $c$, the fitting of $E_F$ and $E_M$ is shown in (c) and $n_S$ in (d). The value of $t_2$ and maximum fitted carrier density greatly depends on $c$. In this case, we found the parameters of $a = 1$, $b = 40$, $t_0 = 1$ and $c = 0.0038$ can fit our experiment data quite nicely.

In the formula $E_F = a/t$, the constant $a$ is determined by the fixed total number of carriers assuming a constant density of states, as expected for a two-dimensional free-electron gas.

In the formula $E_M = b\exp(-t/t_0) + c$, the constant $b$ is evaluated by the strength interface disorder on localizing carriers, and the constant $t_0$ is the screening length. As shown in Fig. S10(a), if b is large and t₀ is small, this means interface defects can strongly localize the electrons (higher position of $E_M$) when STO layer is thin, and this localization effect decays quickly when increasing the STO layer number. If b is small and $t_0$ is large, this localization effect is relative weaker (lower position of $E_M$) but can affect more STO layers. Therefore, these two parameters describe how *interface* defects influence carrier density.

The constant *c* is used to describe the localized electronic states caused by defects which are assumed to be homogeneously distributed in STO layers.

In our calculation, for *a*, *b* and *c*, their relative values are more important than the absolute values, since the value of $E_M/E_F$ determines the carrier density. Apparently, *b* should be largest since the sample with thinnest STO layer shows strong localization at low temperatures. *c* should be smallest, because only the weak localization is observed in the samples with STO layers from 8 to 16 uc. Hence, fixing $a = 1$, we can use $b = 40$ and $c = 0.0038$ to fit our experimental data. The screening length $t_0$, normally it should be 1-2 uc when the interface is sharp. So, we use $t_0 = 1$ in our calculation, which is physically reasonable.